\documentclass{article}

\usepackage{amsmath}
\usepackage{graphicx}
\usepackage{wasysym}
\usepackage{xcolor}
\usepackage{wrapfig}
\usepackage{color}
\usepackage[margin=2cm]{geometry}
\usepackage{multicol}
\usepackage{caption}
\usepackage{authblk}
\usepackage{ragged2e}
\newcommand{\be}{\begin{equation}}
\newcommand{\ee}{\end{equation}}

\author[1]{Steve Markham}
\author[1]{Dave Stevenson}
\affil[1]{California Institute of Technology, Dept. of Geological and Planetary Sciences}
\title{Excitation Mechanisms for Jovian Seismic Modes}

\begin{document}
\maketitle

\begin{abstract}
Recent (2011) results from the Nice Observatory indicate the existence of global seismic modes on Jupiter in the frequency range between 0.7 and 1.5mHz with amplitudes of tens of cm/s. Currently, the driving force behind these modes is a mystery; the measured amplitudes are many orders of magnitude larger than anticipated based on theory analogous to helioseismology (that is, turbulent convection as a source of stochastic excitation). One of the most promising hypotheses is that these modes are driven by Jovian storms. This work constructs a framework to analytically model the expected equilibrium normal mode amplitudes arising from convective columns in storms. We also place rough constraints on Jupiter's seismic modal quality factor. Using this model, neither meteor strikes, turbulent convection, nor water storms can feasibly excite the order of magnitude of observed amplitudes. Next we speculate about the potential role of rock storms deeper in Jupiter's atmosphere, because the rock storms' expected energy scales make them promising candidates to be the chief source of excitation for Jovian seismic modes, based on simple scaling arguments. We also suggest some general trends in the expected partition of energy between different frequency modes. Finally we supply some commentary on potential applications to gravity, Juno, Cassini and Saturn, and future missions to Uranus and Neptune. 
\end{abstract}

\section{Introduction} \label{intro}
Jupiter is the largest planet in the solar system, and our most accurate nearby representation of thousands of exoplanet analogues which seem to be equally or more massive, and comprised of approximately the same material. 
Understanding Jupiter's formation history, then, is of great importance for understanding how planetary systems form in general. 
Understanding Jupiter's interior is an essential part of modeling mechanisms for its formation; for example, the most popular explanation for Jupiter's formation would suggest that the embryo Jupiter was a rocky planet early in its formation history, and we can perhaps expect a many Earth mass core to exist as a relic of that time \cite{core}.
Additionally, there is an abundance of information about thermodynamics and materials physics to be learned by probing the detailed structure of Jupiter's deep interior. 
Current methods of constraining Jupiter's interior (e.g., gravity and magnetic field measurements) are valuable, but cannot uniquely determine the internal structure. 
Therefore seismology will be an indispensable tool as we continue to try to study Jupiter's interior \cite{giant}.
Techniques applied to Jupiter can also be generalized to other planetary systems, and the scientific community has already expressed interest in applying similar techniques to Uranus, Neptune \cite{odinus}\cite{ice-giants}, and even Venus \cite{venus}\cite{logonne}. \\

In 2011, a team from the Nice Observatory released a paper which claimed to have detected normal modes from Jupiter using an interferometer called SYMPA to perform Fourier transform spectroscopy \cite{sympa1}\cite{sympa2}\cite{french}. 
SYMPA measures line of sight Doppler shifts, so the detected displacements are primarily radial. 
For modes within the frequency range of sensitivity (high order p-mode overtones with frequencies above about 700$\mu$Hz), SYMPA detected peak oscillation velocities on the order of 50cm/s. 
As outlined in Section \ref{sph-harm}, this value is the result of the superposition of multiple modes, and the velocity amplitudes of individual modes may be lower by a factor of 2 or 3. 
To put this is perspective, compare this to the maximum velocity amplitude in any single mode found in the sun, around 15cm/s \cite{lectures}. 
The total peak velocities measured on the sun can be substantially higher, because the solar observatory's exquisite spatial resolution allows them to resolve much higher spherical order modes, and therefore more of an effect from superposition. 
Apparently the surface velocity amplitudes of both bodies are of similar orders of magnitude. 
It should be noted that since SYMPA's measurements were limited to eight nights without continuous observations, and because the instrument has low spatial resolution, that these measurements are only relevant to low spherical order, high frequency modes (overtones of global scale modes). 
The power spectrum for the SYMPA measurements is found on Figure~\ref{power-spect}. 

{
\label{power-spect}
\includegraphics[scale=0.6]{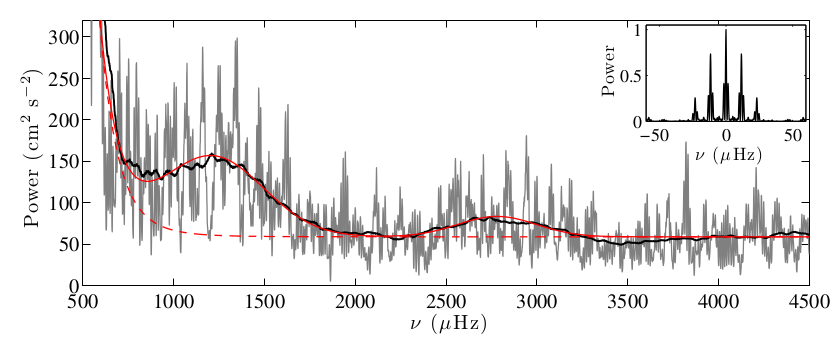}
\captionof{figure}{The observed power spectrum obtained by Gaulme et. al. \cite{french}}
}

This result is encouraging because it means the signal is sufficiently strong that meaningful measurements can be taken from Earth. 
It is puzzling, however, because it requires an excitation mechanism on Jupiter that is fundamentally different from what happens in the sun. 
We can conduct a simple order of magnitude calculation to enumerate the problem here. 
Since each normal mode behaves as a simple harmonic oscillator, its total energy is equal to its maximum kinetic energy. 
If its eigenfunction is described by displacement vector eigenfunction $\mathbf{\xi}$ (further discussed in Section~\ref{model-eigs} and illustrated in Figure~\ref{eig-plots}) normalized to a magnitude of unity at the surface, then integrating over the whole body yields the total energy contained within a given normal mode. 
\be
E_{\text{mode}} = \frac{1}{2} v^2 \iiint \rho |\xi|^2 dV
\ee
where $v$ is the velocity amplitude, $\rho$ is the spatially dependent density. 
$\iiint \rho |\xi|^2 dV$ is called the \textit{modal mass} \cite{lectures}. 
The order of magnitude behavior of the eigenfunctions in the sun and in Jupiter should be similar, so we can neglect that factor since it is not a significant distinction between Jupiter and the sun. 
That is, for similar eigenfunction structure $\xi$, one can approximate the modal mass $\iiint \rho|\xi|^2 dV \sim f M$ to zeroth order--that is, the modal mass scales approximately linearly with the mass of the body \cite{lectures}. 
We can therefore derive a zeroth order scaling relation of the form 
\be 
E_{\text{mode}}\sim M v^2
\ee
where $M$ is the mass of the body. 
Of course, this simplistic analysis ignores relevant details. 
The density contrast between the shallow and deep parts of the sun is much more extreme than for Jupiter; this affects both the modal mass and the excitation efficiency. 
Still, as a zeroth order first approximation to introduce the problem, we can place an order of magnitude estimate on the efficiency with which energy is injected into this normal mode by comparing the squared velocity amplitude to the luminosity per unit mass. 
The luminosity per unit mass in the sun is about 2~erg~g$^{-1}$s$^{-1}$, and for Jupiter it's about $2\times 10^{-6}$~erg~g$^{-1}$s$^{-1}$ \cite{daves-book}. 
The problem then becomes immediately apparent. 
In order to produce the observed normal modes on Jupiter, the mechanism for injecting energy into the modes and retaining energy within the modes must be millions of times more efficient on Jupiter than on the sun. 
This excitation is computed in more detail in Section~\ref{stochast}. 
At the moment, this disparity is not understood. 
The focus of this paper is to attempt to identify mechanisms which could deposit energy into Jupiter's normal modes orders of magnitude more efficiently than the sun. \\

Helioseismology revolutionized our understanding of the sun. 
Studying the sun's seismic modes definitively answered questions ranging from the solar neutrino problem, the sun's convective and radiative zones, the existence of deep jet streams, the age of the sun, and its differential rotation \cite{helioseismology}. 
Today, many fundamental questions about Jupiter may be answered with the same treatment. 
Dioseismology (an alternative word with equivalent meaning to Jovian seismology, first used by Mosser \cite{dios}) could illuminate a condensed or diffuse core. 
It could provide more detailed information about the physical properties of liquid metallic hydrogen, and reveal the existence of regions of static stability or exotic chemical cloud decks deep below the visible surface. 
With so much to gain from dioseismology, it is a worthwhile endeavor to understand. \\

Unfortunately, the existing data for normal modes has rather low signal to noise ratio and is regarded by some as suspect, in part because we lack an understanding of how the modes could be excited. 
If we can develop a more quantitative understanding of their excitation and dissipation, then we could corroborate the possibility of their existence and motivate future observational programs.
Such insights would be useful diagnostic tools to design space-based seismometers for future missions to Jupiter, as well as other planets in the solar system. \\

The 1994 comet strike of Shoemaker-Levy sparked much interest into the possibility of Jovian seismic mode excitation by the cometary impact. 
Competing calculations made contradictory predictions at the time. 
Dombard \& Boughn did not predict measurable amplitudes \cite{dandb}, but others such as Lognonne, Mosser and Dahlen predicted measurable amplitudes for a sufficiently energetic impact \cite{lognonne1994}. 
As it turns out, the seismic modes associated with SL9 were never detected \cite{sl9-dud}. 
In this work, we generalize the framework constructed by Dombard and Baughn for the expected seismic response to the impact of Shoemaker-Levy with Jupiter \cite{dandb}, as well as the work for the sun and other stars made by Peter Goldreich and others \cite{turbulent-excitation-prime}\cite{turbulent-excitation}, to try to propose any plausible candidates for Jovian seismic mode excitations. 
These mechanisms should be both explanatory and predictive; if a certain model explains the observed results, it can also predict what amplitudes should be expected in frequency ranges which have not yet been detected. 
Future measurements, then, can provide support or refutation for different models proposed here. \\

This paper will begin with an introduction to our model of Jupiter and the treatment of its normal mode displacement eigenfunctions. 
We will then outline some general mathematical tools to abstractly model and parameterize different types of excitation sources. 
Next we will investigate a few important dissipation mechanisms to try to place some constraints on Jupiter's modal $Q$. 
We will then apply all these tools to some potential physical excitation sources, to try and estimate an order of magnitude for what velocity amplitudes these mechanisms might excite. 
Finally we will discuss our findings, with some brief remarks on potential applications of these findings to Jupiter and other planets. 

\section{Modeling the eigenfunctions of Jupiter's seismic modes} \label{model-eigs}
Jupiter, like any other object, can behave as a resonator. 
The modes of interest for explaining the results from SYMPA are acoustic modes. 
These modes are trapped in a cavity bounded from below by Snell's law; the ray path enters Jupiter's interior from the surface obliquely. 
As the ray descends, the sound speed increases, which continuously deflects the ray laterally until it travels tangentially at the minimum radius and begins to return to the surface. 
Modes below the acoustic cutoff frequency are bounded from above by Jupiter's small scale height (relative to the mode's local wavelength) as it approaches the photosphere. 
This resonator is rather efficient, since the viscosity in Jupiter is very low. 
Much work on this basic physics has been done, primarily with applications to helioseismology and asteroseismology in general \cite{lectures}. 
There has also been some qualitative work on applying these ideas to Jupiter \cite{bs}. 
Some progress can be made by qualitative order of magnitude arguments, but in order to argue for a coherent global picture, a numerical model for the structure of the eigenfunctions, the planetary interior, and the planetary atmosphere must be specified. 

\subsection{Jupiter Interior Model}
The first important step in this modeling process is choosing a suitable Jupiter interior model. 
This model can in principle be as detailed as desired, but for our purposes we wanted to use the simplest, most generic possible model that can still accurately model Jupiter's behavior because our focus here is on understanding the excitation and dissipation, not the precise evaluation of modal eigenfrequencies. 
This is desired for simplicity of outcome (no frequency splitting between modes of the same spherical order), as well as simplicity of inputs (homogeneous adiabatic interior), and finally for its ability to easily adapt to explain other planets. 
We therefore begin with a simple $n=1$ polytrope equation of state 
\be P=K\rho^2 \ee
with $K$ chosen to approximate a hydrogen/helium mixture. 
This model is quite accurate for Jupiter's interior, but does a bad job at accurately describing the behavior near the surface. 
We therefore adjust the equation of state by adding a $\rho^{1.45}$ term consistent with an adiabatic ideal gas equation of state. 
The two should connect smoothly in between. 
The equation of state then takes the form 
\be P=K_1\rho^2+K_2\rho^{1.45} \ee
where $K_1$ and $K_2$ are chosen to match Galileo measurements for Jupiter's upper troposphere, and to get the right radius and mass. \\
{
\centering
\includegraphics[scale=0.8]{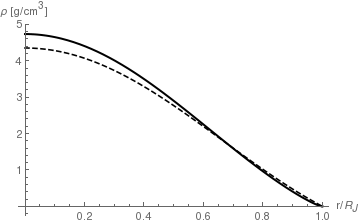}
\captionof{figure}{Comparison between the hydrostatic interior model using our modified equation of state (solid) and the interior model predicted using an n=1 polytrope equation of state (dashed).}
\label{interior-model}
}

Notice that since the $\rho^2$ term is small near the surface, the ideal gas term will then dominate. 
Additionally, we investigated the effects to the eigenfunction if we include an isothermal component to the atmosphere above the photosphere. 
We found that doing so affected observed mode amplitudes by less than 5\%. 
Since the arguments we are making here are generic and correct to no more than an order of magnitude, we elected to neglect the isothermal part of the atmosphere for the purpose of generating the global eigenfunctions. 
We do, however, discuss the effects of radiative damping in the isothermal part of the atmosphere as it relates to Jupiter's quality factor in Section \ref{constrainQ}. 

\subsection{Displacement Vector Eigenfunction Generation}
After setting upon an interior model which satisfactorily represents the important aspects of Jupiter's interior, we used the stellar oscillation code GYRE \cite{gyre} to generate eigenfunctions for Jupiter's interior. 
The first four l=2 modes are shown on Figure~\ref{eig-plots}\\
{
\centering
\includegraphics[scale=0.7]{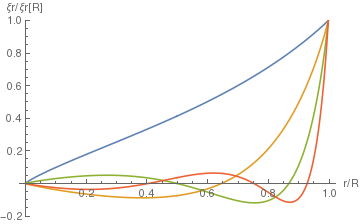}
\captionof{figure}{An example of the radial eigenfunction produced for our interior model the first four $l=2$ modes. $\xi$ represents the amplitude of the eigenfunction in the radial direction at that depth, normalized such that $\xi=1$ at the 1 bar level.}
\label{eig-plots}
}
Because we are using a non-rotating, spherically symmetric model for Jupiter, the modes are exactly spherical harmonics. 
The behavior of the eigenfrequencies is shown on Figure~\ref{freqs}. \\\\
{
\centering
\includegraphics[scale=0.7]{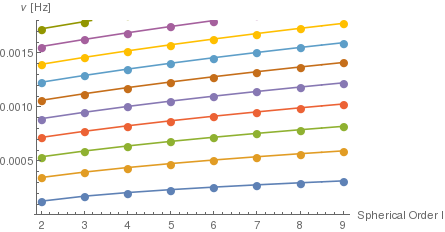}
\captionof{figure}{Frequencies of low order modes. Frequency increases gradually with increasing spherical order $l$ and quickly with equal spacing with increasing radial order $n$, where $n$ defines the number of nodes of the mode as shown in Figure~\ref{eig-plots}.}
\label{freqs}
}
The total observed displacement on the surface of Jupiter is expressed as 
\begin{equation}
\mathbf{x}(\mathbf{r},t)=\sum_{nlm} a_{nlm}(t)\mathbf{\xi_{nlm}}(\mathbf{r})
\label{superposition}
\end{equation}
where $a_{nlm}(t)$ is a time dependent amplitude for each normal mode, and $\mathbf{\xi_{nlm}}(\mathbf{r})$ is a spatially dependent eigenfunction displacement vector of radial order $n$, spherical order $l$ and azimuthal order $m$. 
Canonically the eigenfunctions are separated into a radial and horizontal part $\xi_r(r)$ and $\xi_h(r)$ so that the full displacement vector eigenfunction takes the form 
\begin{equation}
\mathbf{\xi_{nlm}}(r,\theta,\phi)=\left[ \xi_r(r)\mathbf{\hat{r}}+\xi_h(r)\mathbf{\hat{\theta}}\frac{\partial}{\partial\theta}+\frac{\xi_h(r)}{\sin\theta}\mathbf{\hat{\phi}}\frac{\partial}{\partial\phi}\right] Y_l^m(\theta,\phi)
\label{eigs}
\end{equation}

\section{Modeling and parameterizing amplitude responses from generic local excitation sources}
For the purposes of this problem, we will approximate the modes of Jupiter as a set of orthogonal, undamped harmonic oscillators.
This is a valid approximation because our assumed timescales for damping is proportional to a very large $Q$. 
Specifically, $\tau_{dec}=2 Q/\omega$ for a given mode, and we expect $Q$ to be $\sim 10^6-10^8$, which we will justify later in this paper. 
Since $Q$ is so large, we will approximate the timescale between excitation events to be much less than the ringdown timescale. 
As another approximation, we will assume no ``leaking'' energy between modes, i.e., the modes are linear and non-linear interaction terms are neglected, but this will be discussed in our evaluation of $Q$. 
Now we write down the equation of a driven harmonic oscillator 
\be \ddot{a}+\omega^2 a=F(t) \ee
for each mode, where $a$ is the time dependent coefficient from Equation~\ref{superposition}, $\omega$ is the appropriate eigenfrequency, and $F(t)$ is an effective force. 
For a mass on a spring, this effective force would simply be the physical force divided by the mass of the object. In this simplified case, the whole driving force acts on the whole mass, but since our excitation sources may be localized, we must define the effective force following Dombard \& Boughn \cite{dandb}. 
This effective force should account for the coupling between the eigenfunction $\xi$ and the physical force density vector field $\mathbf{f}(\mathbf{r},t)$, and scale it by the total modal mass. 
\begin{equation}
F(t)=\frac{\iiint \mathbf{\xi}\cdot\mathbf{f} dV}{\iiint \rho(\mathbf{r})|\mathbf{\xi}|^2 dV}
\label{force}
\end{equation}
In the following subsections, a few simple generic models for force density will be examined. 
Later in the paper, these generic models can be combined to approximately model physical phenomena to an order of magnitude. 

\subsection{Monopole Excitation}
An explosion is an example of a monopolar force density field. 
Following the model of Dombard \& Boughn \cite{dandb} for a comet impact, we can model a spherical explosion centered on a point $\mathbf{r_0}$ as 
\be \mathbf{f}(\mathbf{r},t)=\delta P \delta(\mathbf{r}-\mathbf{r_0})\mathbf{\hat{r}}_n\phi(t) \ee
where $\delta P$ is the pressure pulse caused by the explosion, $\delta(\mathbf{r}-\mathbf{r_0})$ is a spherical delta function, $\mathbf{\hat{r}}_n$ is an unit vector pointing away from $\mathbf{r_0}$, and $\phi (t)$ is an arbitrary function in time which sets the timescale of the explosion.  
Substituting this $\mathbf{f}$ into Equation~\ref{force}, using Gauss' theorem, and noting that the energy of the bubble is equal to its pressure perturbation times the volume of the bubble gives
\be F(t)=\frac{ E_s/V \phi(t) \iiint_s \nabla \cdot \xi d^3 r}{\iiint \rho(r) |\xi|^2 d^3 r} \ee
where $V$ is the volume of the bubble, $E_s$ is its energy, and the integral in the numerator is over the volume of the explosion. 
Our task is now to compute this expression. 
Assuming that there is very little non-radial variation in $\nabla\cdot \xi$ (which is a very good approximation near the planetary surface for excitation sources with length scales on the order of hundreds of kilometers, as long as we are talking about spherical orders less than several thousand) we can simplify 
\be \iiint_s \nabla\cdot\xi d^3r \rightarrow \pi \int_{-b}^b \nabla\cdot\xi (x^2-b^2) dx \ee
where $b$ is the radius of the bubble. 
We can do a Taylor series $\nabla\cdot\xi$ up to a fourth derivative in $\xi_r$, which is more than a good enough approximation for these length scales with $n < 100$, we can compute this integral directly to be 
\be \iiint_s \nabla\cdot\xi d^3r \approx 4/3 \pi b^3 \left(\frac{\partial\xi_r}{\partial r} + 1/10 \pi b^2 \frac{\partial^3\xi_r}{\partial r^3}\right) 
\label{thatone}
\ee
since the displacement eigenfunction for low spherical order $l$ modes is primarily radial near the surface. 
This approximation breaks down for higher spherical order modes, where the tangential component of the eigenfunction is more important. 
To ensure the accuracy of this method, we compared the exact numerical integration of the divergence of the eigenfunction through the bubble to this approximation, and found excellent agreement for the first fifty modes within less than 1\%. 
In fact, for the first 25 modes (which are the ones in the frequency range of interest), the third order term in also unnecessary. 
Since $4/3\pi b^3$ is constant, we can take it out of the integral. 
It's also the volume of the bubble, so we can cancel it with $V$. 
Thus if we approximate the spatial and time dependence to be separable quantities, we can write  
\be F(t)\simeq \frac{E_s \frac{\partial\xi_r}{\partial r} }{\iiint \rho(r)|\xi|^2d^3r}\phi(t)\equiv F_0\phi(t) \ee
For high radial order modes ($n>30$), the $\frac{\partial^3 \xi_r}{\partial r^3}$ term from Equation~\ref{thatone} should be included for accuracy. 
Now we can solve the harmonic oscillator equation 
\be \ddot{a}+\omega^2 a=F(t)=F_0\phi(t) \ee
where $F_0$ encodes the geometric information, assumed spatially static in space and wrapped in a time dependent wrapper function $\phi (t)$. 
Since $F_0$ is a constant in time, in the one dimensional harmonic oscillator equation it can be considered to be a constant. 
We now solve this equation by taking its Fourier transform, so that 
\begin{equation}
a(t)=\frac{F_0}{\sqrt{2\pi}}\int_{-\infty}^\infty \frac{\hat{\Phi}(\nu)}{\omega^2-\nu^2}e^{i\nu t}d\nu
\label{fourier}
\end{equation}
where $\hat{\Phi}(\nu)$ is the Fourier dual of $\phi(t)$. 
All that is required, then, is to choose a form of $\phi(t)$ and Equation~\ref{fourier} is solvable.

\subsection{Dipole Excitation}
The simplest way to think of a dipole is two point sources separated by some distance $\epsilon$. 
This is expressed mathematically as 
\be \mathbf{f}(\mathbf{r},t)=f_0[-\delta(\mathbf{r}-(\mathbf{r_0}+\epsilon))+\delta(\mathbf{r}-\mathbf{r_0})]\hat{\mathbf{r}} \ee
where $\hat{\mathbf{r}}$ is the outward pointing radial vector with respect to the center of Jupiter, and $f_0$ is the normalization coefficient. 
Provided $\epsilon$ is small compared to the wavelength of the mode, a reasonable first order approximation, we can evaluate 
$$\iiint \mathbf{f}(\mathbf{r})\cdot \mathbf{\xi} dV \approx -f_0 \frac{\partial \xi_r}{\partial r}\epsilon$$
\begin{equation}
\implies F_{0,dipole}\sim \frac{f_0 \frac{\partial \xi_r}{\partial r} \epsilon}{\iiint \rho(r)|\xi|^2 dV}
\label{dipole}
\end{equation}
using the fundamental theorem of calculus and the properties of the $\delta$ function. 
For our purposes, this is a sufficient description of a generic dipole excitation. 
For a specific model, of course, one must evaluate a physically reasonable $f_0$ in the context of the problem. 
Note the striking similarity between localized dipole and monopole excitation sources, which for low spherical and radial order modes are mathematically identical, except with different expressions for $F_0$.

\subsection{Spatial Randomness}
In all of the above results, the predicted amplitudes implicitly include a spherical harmonic evaluated at a particular point on Jupiter's surface. 
If at any instant there are $N$ storms within Jupiter's atmosphere then the total displacement would scale as 
\be \sum_{i=1}^N |Y_l^m (\theta_i,\phi_i)|^2 = N \ee
In the limit of large $N$ and assuming the storms are randomly distributed, the RMS value of this is simply $N^{1/2}$ larger than the ampltiude of a single storm, because of the normalization properties of spherical harmonics. 
Of course this would break down in the limit of small number of storms, or storms with a preferred location, as may be the case. 
In this case, there may be more complicated dependence of amplitude on the quantum numbers than the results we report below.

\subsection{Temporal Randomness}
Having shown that spatial randomness of storm occurrence can be averaged out to be irrelevant, the next logical question is what to do about the issue of the storms being stochastic in time. 
Because of the findings in the previous section, geometrical effects can be neglected. 
The amplitude response from a single excitation event $j$ takes the form 
\be \mathbf{x}_j(\mathbf{r},t)=\sum_{nlm} a_{nlm,j} \mathbf{\xi}_{nlm} \exp(i \omega_{nlm}(t-t_j)) \ee
The full expression after $N$ excitations can be written 
\be \mathbf{x}(\mathbf{r},t)=\sum_{nlm}\sum_{j=1}^N [a_{nlm,j} \exp(-i \omega_{nlm} t_j)] \mathbf{\xi}_{nlm} \exp(i \omega_{nlm} t) \ee
The task now is to evaluate 
\be \sum_{j=1}^N a_{nlm,j}\exp(-i \omega_{nlm} t_j) \ee
since $t_j$ is a random variable, and $\exp(-i \omega_{nlm} t_j)$ is a $2\pi$ periodic function, the above expression is simply a random walk in the complex plane. 
The final expression for the amplitude without dissipation after $N$ excitation events then can be written 
\be \mathbf{x}(\mathbf{r},t)\approx \sqrt{N} \sum_{nlm} a_{nlm,j} \mathbf{\xi}_{nlm} \cos(\omega_{nlm} t+\phi) \ee
where $\phi$ is an arbitrary phase and $a_{nlm,j}$ is now the expected value of amplitude for a given type of excitation. 
Because the energy of the mode scales as $|x|^2$, 
energy grows linearly with the number of excitation events, while amplitude grows with its square root. \\

Now we calculate the equilibrium mode amplitudes including dissipation. 
If a single excitation imparts energy $E_0$, and the expected value for total energy input grows linearly with the number of excitation events, then we can equate average power input to energy dissipation 
\be \frac{E_0}{\tau_{s}}=\frac{E_{eq}}{\tau_{dec}} \ee
where $\tau_{s}$ is the characteristic timescale between excitation events, and $\tau_{dec}$ is the decay timescale, related to the quality factor $Q$ according to 
\be \tau_{dec}=\frac{2 Q}{\omega} \ee
Of course this assumes that there is an equilibrium i.e., the time between excitation events is much shorter than the time to decay. 
If this were not so, it would be evident in continued observations that show a variation of mean amplitude over time. 
The mean equilibrium energy associated with an excitation source that imparts energy $E_0$ stochastically in time is 
\begin{equation}
E_{eq}=\frac{2 E_0 Q}{\tau_{s} \omega}
\label{eqilib}
\end{equation} 
It should be noted that these values are not expected to be constant in time. 
The arguments here are only statements about the average equilibrium amplitudes; in reality, one observes a specific amplitude at a specific time rather than a long term average. 
It is therefore perfectly consistent with this framework to have periods of quiescence, and periods of larger amplitudes. 
The expected value, however, will tend toward the calculations shown here. \\

As argued in Section~\ref{intro}, the energy of a mode described by displacement eigenfunction $\xi$ is 
\be E_0 = \frac{1}{2} a^2\omega^2 \iiint \rho |\xi|^2 dV \ee
where $a$ is the amplitude response resulting from a single excitation. 
Ignoring time dependence and focusing on amplitude, we can use $a=F_0/\omega^2$. 
In reality, the form of $a$ will depend on $\phi(t)$, but that's the focus of the following section. 
Rewriting $E_0$ as  
\be E_0=\frac{1}{2} F_0^2 \omega^{-2} \iiint \rho |\xi|^2 dV \ee
The equilibrium amplitude is 
\be a_{eq}=\left(\frac{E_{eq}}{\omega^2 \iiint \rho|\xi|^2 dV}\right)^{1/2} \ee
so using Equation~\ref{eqilib}, the equilibrium amplitude can be written 
\begin{equation}
a_{eq}=F_0\left( \frac{Q}{\tau_{s} \omega^5} \right)^{1/2}
\label{fundamental}
\end{equation}
This relation is of enormous consequence for Jovian seismic mode excitation. 
The forcing magnitude of a generic source is proportional to its energy scale. 
Equation~\ref{fundamental} implies that the equilibrium amplitude obeys 
\be 
a_{eq}\propto \frac{E_s}{\tau_{ex}^{1/2}}
\ee
While the power output of these collective excitation sources by definition follows the relationship 
\be 
\dot{E}= \frac{E_s}{\tau_{ex}}
\ee
\textit{Hence, for a fixed power budget, it is more favorable to have less frequent, more energetic excitation events than more frequent, less energetic excitation events.}

\subsection{Excitation duration}
The dynamics of storms are immensely complex. 
Decades of detailed research have gone into modeling storms on Earth for which we have excellent data, and still there is no basic universal picture for their dynamics \cite{earth-storms}. 
For the purposes of this paper, the time dependent aspect of storms as an excitation source will be modeled simplistically. 
In particular, the Heaviside Theta function, a Gaussian function, and a hat function will be considered. 
Recalling Equation~\ref{fourier}, we can solve for each of these. 
For a Heaviside Theta function, 
\be \phi(t)\rightarrow \Theta(t)\implies a=F_0/\omega^2 \ee
The expectation here is that lower frequencies would receive greater excitation, for constant $F_0$. 
For a Gaussian, 
\be \phi(t)\rightarrow \exp(-t^2/2\Delta t^2) \implies a=\sqrt{2 \pi} \sigma \frac{F_0}{\omega}\exp(-\omega^2\Delta t^2/2) \ee
where $\sigma$ sets the width of the Gaussian and has dimensions of time. 
In this case, the narrower the Gaussian for the input $\phi(t)$, the broader the excitation spectrum in frequency space. 
For a hat function, 
\be \phi(t)\rightarrow 1 \text{ for } |t|<\Delta t \text{, 0 elsewhere } \implies a=\frac{2 F_0}{\omega^2}\sin(\omega \Delta t) \ee
Here the amplitudes in frequency space are a sinusoid, so there is no explanation as simple as for the Gaussian for all frequencies. 
However, for frequencies which satisfy $\omega\Delta t\ll\pi/2$, the same basic principle applies. 
The narrower the hat function, the broader its excitation in frequency space. 
Note we have not investigated the delta function here. 
This is because the delta function is not dimensionless, and therefore cannot be used for this purpose. 
For our storm models, our choice of excitation duration timescale, $\Delta t$, will 
impact on the results.

\subsection{Spherical harmonic superposition in the power spectrum} \label{sph-harm}
So far these calculations have focused on the excitation of a single mode given some source. 
This section remarks briefly on the expected power spectrum that would be measured from all visible modes combined. 
We begin with the general mathematical relationship 
\be \int_{\phi=0}^{2\pi}\int_{\theta=0}^{\pi}\left| \sum_{l=0}^N\sum_{m=-l}^l Y_l^m \right|^2 \sin\theta d\theta d\phi=N \label{spherical-harmonic-superposition}\ee
Recall that the expression for excitation amplitude given the sources investigated here depend on $\frac{\partial \xi_r}{\partial r}$ and $\omega$. 
The only dependence on $Y_l^m$ is encoded in the denominator, since 
\be |\xi|^2=\mathbf{\xi}\cdot\mathbf{\xi}=\xi_r^2 \left|Y_l^m\right|^2 + \xi_h^2 \left|\frac{\partial Y_l^m}{\partial \theta} \right|^2 \ee
This expression is integrated over a sphere, so the $\left|Y_l^m\right|^2$ averages away. 
The $\left|\frac{\partial Y_l^m}{\partial \theta} \right|^2$ is retained, but for sufficiently low order spherical harmonics near the surface,
the motions are mostly radial, so the second term can be neglected. 
Since $\xi_r$ 
is independent of $m$ and only weakly dependent on $l$ for low spherical order modes, this implies that to a good approximation the excitation amplitude is a function of frequency only. 
This means that assuming SYMPA is sensitive to spherical orders up to about $l=3$, the power spectrum calculated for one spherical mode can be approximately doubled to account for the full power spectrum. 
On the sun, where resolution is greatly enhanced and detection of very high spherical order modes are possible, we expect this principle to have a more substantial effect on peak measured velocity, because the higher resolution implies detection of higher $l$ modes and therefore larger $N$ in Equation~\ref{spherical-harmonic-superposition}. 

\section{Constraining Q} \label{constrainQ}
As demonstrated in the previous section, our equilibrium mode amplitudes scale as $Q^{1/2}$. 
Having an idea for the order of magnitude of Jupiter's quality factor, then, is essential to making a predictive theory. 
One possibility is that the effective $Q$ is actually determined by the interaction of modes with each other rather than intrinsic dissipation. 
However, these interactions are probably negligible \cite{jing}, so for the moment we will focus on intrinsic processes. 
Much work has already been done estimating Jupiter's tidal $Q$ \cite{inertial-modes2}. 
The primary coupling mechanism between Jupiter and its satellites are inertial modes, which are bounded between $0<\omega<2 \Omega$, where $\Omega$ is Jupiter's spin rate \cite{inertial-modes}\cite{inertial-modes2}. 
The fundamental p-mode of Jupiter has a period on the order of two hours, much shorter than Jupiter's spin rate. 
Therefore dissipation associated with these inertial modes is irrelevant to the study at hand. 
Nevertheless, it is possible to place some constraints on our expected value of $Q$ using mechanisms we know must dissipate energy. 

\subsection{Viscous and turbulent damping}
The most obvious dissipation mechanism is viscosity. 
Starting with the standard Stokes-Kirchhof viscous dissipation expression for acoustic waves \cite{landl}, 
\begin{equation}
\bar{\dot{E}}=-\frac{1}{2}k^2 v_0^2 V_0 \left[ \left( \frac{4}{3}\eta+\zeta \right)+\kappa \left(\frac{1}{c_V} - \frac{1}{c_p} \right) \right]
\label{viscosity}
\end{equation}
where $k$ is the sound wavenumber, $v_0$ is the fluid displacement velocity, $V_0$ is the volume occupied by the sound wave, $\eta$ is dynamic viscosity, $\zeta$ is the second viscosity, $\kappa$ is the fluid's thermal conductivity, $c_V$ is the specific heat capacity of the fluid at constant volume and $c_p$ is the specific heat capacity at constant pressure. 
As a simplifying assumption, assume $\zeta\sim\eta$. 
Now compare the relative importance of the the first and second bracketed terms on the right hand side of Equation~\ref{viscosity}. 
Noting $\kappa (1/c_V-1/c_p) = \kappa/c_p(\gamma-1)$ and plugging in typical values for hydrogen, the second term is $\sim 10^{-12}$ in cgs units, compared to viscosity which is $\sim 10^{-3}$. 
So the second term can be neglected. 
Now we write 
\be \bar{\dot{E}} \approx -k^2 \omega^2 |\xi|^2 V_0 \eta \ee
Integrating over differential volume elements, we get a total average power dissipation of 
\be \bar{\dot{E}} = \omega^2 \iiint k^2 \eta |\xi|^2 dV \ee
Now to compute $Q$, note
\be Q \equiv 2 \pi \frac{E_{\text{stored}}}{\oint \bar{\dot{E}} dt}=\omega \frac{\iiint \rho |\xi|^2 dV}{\iiint k^2 \eta |\xi|^2 dV} \ee
Now for order of magnitude estimates, assume $k$ to be constant to zeroth order in most of the interior. 
Substitute average, constant values $\bar{\rho}$ and $\bar{\eta}$ and take them out of the integral. 
The expression for $Q$ then reduces to 
\be Q \sim \frac{\omega}{k^2} \frac{\bar{\rho}}{\bar{\eta}} \ee
Noting $k\sim \frac{2\pi (n+1)}{R_{\jupiter}}$ where $n$ is the radial order, and $\bar{\rho} \approx 1.33$. 
This gives 
\be Q\sim 10^{18} \left( \frac{\omega}{10^{-3}s^{-1}} \right) \left(\frac{1}{n_r+1} \right)^2 \left(\frac{10^{-2}\text{cm}^2\text{s}^{-1}}{\eta}\right) \ee
where $n_r$ is the radial order of the mode. 
In reality, turbulence will increase the effective viscosity of the system. 
Turbulent viscosity should be weak, because Jupiter's convection overturn timescale is much longer than the period of the normal modes, which means eddies larger than the local scale height do not act viscously \cite{viscosity}. 
Assuming $\eta\sim 10^3$ as is assumed for tides \cite{viscosity}, the estimate for $Q$ goes to $\sim 10^{13}$. 
So viscosity and turbulence turn out to be very weak damping mechanisms.

\subsection{Radiative damping}
The most important mandatory loss of energy occurs as a result of radiative damping in Jupiter's stratosphere. 
Below the tropopause, a displaced parcel of fluid will expand or contract adiabatically, but remain in equilibrium with its convective surroundings, which by definition follow an adiabat. 
However, the same displacement in the isothermal atmosphere would cause a displaced parcel to warm as it was displaced downward, bringing it out of equilibrium with its surroundings. 
The warm parcel would then radiate away heat while displaced. 
Conversely, a parcel displaced upwards will radiate less heat. 
Importantly, this introduces a phase difference between the oscillations in temperature associated with a wave and the oscillations in pressure or density. 
The resulting hysteresis is the dissipation arising from radiative damping. 
We are primarily interested in the case where the tropopause occurs at a location where the waves of interest are no longer propagating (i.e., are evanescent) so that the effect of the wave on the atmosphere is merely the vertical displacement of a column of gas. 
In the low frequency limit, the fractional density perturbation and the velocity amplitude increases only slightly with height, with a characteristic e-folding distance of $c^2/\omega^2 H \sim R_{\jupiter}$. \\

First we calculate the radiative damping timescale $\tau_{rad}$. 
Assuming the atmosphere is optically thin in the stratosphere, and gray opacity such that emission and absorption are described by the same constant, we imagine a parcel in an isothermal environment of temperature $T_0$ raised to temperature $T_0+T'$ by being displaced by seismic modes. 
It is illuminated from below by the ammonia cloud deck of optical depth unity at Jupiter's effective temperature $T_e$. 
The total energy radiated from the plane parcel up and down is 
\be 2\sigma (T_0+T')^4\rho\kappa dz \ee
energy absorbed from below is 
\be \sigma T_e^4\rho \kappa dz \ee
In equilibrium with $T'\rightarrow 0$, we obtain the standard result $T_0=T_e/2^{1/4}$. 
On the other hand, out of equilibrium with time dependent $T'$:
\be \rho c_p dz \frac{d T'}{dt}=-8 \sigma T_0^3 T' \rho \kappa dz \ee 
We can write $T_0$ in terms of $T_e$ from the standard result, so that $8\sigma T_0^3 \rightarrow 4\sigma T_e^3$. 
Now defining a radiative time constant $\tau_{rad}$ according to 
\be -\frac{T'}{\tau_{rad}} = \frac{dT'}{dt} \ee 
reveals 
\be \tau_{rad}=\frac{c_p}{4\sigma T_e^3 \kappa} \ee
using values from Galileo, and employing a functional form of pressure dependent opacity for hydrogen as 
\be \kappa\sim 10^{-2}\left(\frac{p}{\text{1 bar}}\right)\text{cm}^2/\text{g} \ee
yields 
\be \tau_{rad}\approx 5\times 10^7 \left(\frac{\text{1 bar}}{p}\right)\text{sec} \ee
Now to calculate dissipation. 
Starting with the ideal gas law 
\be p=\frac{k_B}{\mu}\rho T \ee
\be \implies dp=\frac{k_B}{\mu}(d\rho T + \rho dT) \ee
We are interested in the part of the pressure perturbation associated with the change in temperature. 
So 
\be \delta p \approx \frac{k_B}{\mu}\rho_0 T' \ee
to first order. 
In general for a displaced parcel 
\be \frac{\partial T'}{\partial t}=-v\left(\frac{\partial T}{\partial z}-\frac{\partial T}{\partial z}\bigg\rvert_{ad}\right)+\frac{T'}{\tau_{rad}} \ee
where $v$ is the local velocity of the parcel caused by normal mode oscillations. 
In the isothermal atmosphere, $\frac{\partial T}{\partial z}\rightarrow 0$. 
In general for a plane-parallel atmosphere, $\frac{\partial T}{\partial z}\rvert_{ad}=g/c_p$. 
So assuming $v$ and $T'$ oscillate with the normal mode and are therefore $\propto \exp(i\omega t)$, we can rewrite 
\be T'=\frac{v g}{c_p(i\omega + 1/\tau_{rad})} \ee
Assuming $\frac{1}{\omega\tau}\ll 1$, true using characteristic values of $\tau\sim 5\times 10^7$s and $\omega\sim 10^{-3}$s$^{-1}$, this can be written as 
\be T'\approx \frac{vg}{i \omega c_p}\left(1-\frac{i}{\omega\tau_{rad}}\right) \ee
Substituting this into the ideal gas equation yields 
\be \delta p \approx \frac{k_B}{\mu} \frac{v}{H c_p i\omega}\left(1-\frac{i}{\omega\tau_{rad}}\right)p_0 \ee
by noting $g=\frac{c_s^2}{\gamma H}$ where $c_s$ is the speed of sound and $\gamma$ is the adiabatic index; and that $p_0=c_s^2 \rho_0/\gamma$. 
The task now is to compute the energy dissipated in one normal mode period. 
\be \oint v \delta p dt = \int_0^{2\pi/\omega} v\delta p dt \ee
Now because the quality factor is defined as 
\be Q\equiv 2\pi \frac{\text{stored energy}}{\text{energy dissipated in a cycle}}=2\pi\frac{\omega^2 a^2\iiint \rho|\xi|^2 dV}{4\pi R_{\jupiter}^2\oint v \delta p dt} \ee
The complex exponential of the temperature perturbation term is
\be \frac{e^{i \omega t}}{i}-\frac{e^{i \omega t}}{\omega \tau_{rad}}=-\left(\sin(\omega t)+\frac{1}{\omega \tau_{rad}}\cos(\omega t)\right) \ee
Using the harmonic addition theorem this can be rewritten as a sinusoid with a coefficient and a phase. 
Again using the fact that $\frac{1}{\omega\tau_{rad}}\ll 1$, we can solve the integral over the period to be 
\be \oint v \delta p dt=\frac{k_B}{\mu}\frac{v^2}{H c_p \omega} c \oint \cos(\omega t)\sin(\omega t+\phi)dt \approx \boxed{\frac{\pi k_B v^2 p_0}{\mu H c_p \omega^3 \tau_{rad}}} \ee
Now computing $Q$ to an order of magnitude, and noting $\iiint\rho dV=M_{\jupiter}$ and thus taking $\iiint\rho|\xi|^2 dV\sim M_{\jupiter}/10$ as an order of magnitude approximation based on the behavior of the eigenfunctions, we can write 
\be Q\sim \frac{\mu H c_p \omega^3 \tau_{rad} M_{\jupiter}}{20 \pi k_B p_0 R_{\jupiter}^2}\sim \boxed{10^7\left(\frac{\omega}{10^{-3}\text{s}^{-1}}\right)^3} \ee
This is an upper bound for $Q$, and only correct to an order of magnitude. 
Since it's the best to go on, we will use $Q\sim 10^7$ throughout this work. 

\subsection{High frequency modes: propagation through the stratosphere}
For modes of frequency above the acoustic cutoff frequency, approximated as 
\be \omega_a = \frac{c_s}{2 H} \ee
for an isothermal atmosphere, the modes behave differently. 
For Jupiter, this corresponds to about 3mHz \cite{mosser1995}\cite{cutoffs}. 
Instead of being trapped in Jupiter's interior, with an evanescent tail in the stratosphere, modes above this cutoff frequency propagate into the atmosphere, and eventually into space, unhindered. 
In this  case, the full power of the waves propagating into the statosphere is lost, not just the part out of quadrature. 
The energy density of the waves are given by 
\be \frac{dE}{dV}\sim \frac{1}{2} \rho v^2 = \frac{1}{4} \rho \omega^2 \xi_r^2 \ee
where the additional factor of 1/2 comes from averaging square velocity over a period (since $\xi_r$ is an amplitude). 
These are acoustic modes, so they propagate at the sound speed $c_s=\sqrt{\frac{\gamma k_B T}{\mu}}$. 
So the energy flux through a unit area is given by 
\be \mathcal{F}\sim \frac{1}{4} \rho \omega^2 \xi_r^2 c_s \ee
The total average power loss then is just $\langle \dot{E} \rangle = 4 \pi R_{\jupiter}^2 \mathcal{F}$. 
Relating this to $Q$, 
\be Q \equiv 2 \pi \frac{E_{\text{stored}}}{\oint \langle \dot{E} \rangle dt} \ee
by definition, $\oint \langle \dot{E} \rangle dt=\frac{2 \pi}{\omega} \langle \dot{E} \rangle$ so 
\be Q = \omega \frac{\iiint \rho |\xi|^2 dV}{\pi R_{\jupiter}^2 \rho \xi_r^2 c_s} \ee
Substituting approximate values gives 
\be \boxed{Q \sim 6 \times 10^3 \frac{\omega}{10^{-3}s^{-1}}} \ee
We will not actually use this value of $Q$, but we do this calculation to demonstrate that we should expect any modes with frequencies above the cutoff frequency should not have significant amplitudes relative to modes below it. 

\subsection{Ohmic Dissipation by normal modes}
From the induction equation 
\be \frac{\partial \mathbf{b}}{\partial t} = -\nabla \times (\lambda \nabla \times \mathbf{b})+\nabla\times(\mathbf{u}\times\mathbf{B}) \ee
where $\mathbf{b}$ is the induced field resulting from the action of the normal mode velocity $\mathbf{u}$ acting on the main planetary field $\mathbf{B}$. 
The magnetic diffusivity is $\lambda$, whose value is small (a metal) deep down but large (a semi-conductor) as one approaches the surface. 
Evidently 
\be |\mathbf{b}|\equiv b \sim \frac{k u B}{i \omega+\lambda k^2} \ee
where $k$ is the characteristic wave vector describing the spatial variation of $\mathbf{b}$. 
the Ohmic dissipation per unit volume is $\frac{\lambda(\nabla\times\mathbf{b})^2}{\mu_0}$ and scales as $1/\lambda$ at large $\lambda$ but as $\lambda$ at small $\lambda$. 
The peak dissipation occurs in the region where $\omega \sim \lambda k^2$. 
Dividing kinetic energy of the wave by the dissipation per wave period, we see that 
\be Q_{\text{Ohmic}}\sim 10 \left( \frac{\omega}{V_A k} \right)^2 \ee
where $V_A$ is the Alfven velocity, $\sqrt{B^2/\rho \mu_0}$. 
The coefficient allows for the fact that the volume of dissipation is much smaller than the entire planet and may be an underestimate depending on the conductivity profile. 
This predicts $Q>10^{10}$ for Jupiter, so we do not expect it to be the dominant dissipation mechanism. 

\subsection{Normal Mode dissipation in the core}
An alternative tidal dissipation mechanism, suggested long ago \cite{dermott} assumes that $Q$ is dominated by the small central core, which dissipates in much the same way as a solid terrestrial planet, but possibly aided by soft rheology \cite{storch} or partial melting. 
In this picture, the intrinsic $Q$ of the core is low but the $Q$ of the planet as a whole is higher by several orders of magnitude, simply because of the quadratic dependence of tidal potential on radius and the smallness of the volume involved. 
For modes of spherical order greater than zero, the core is also expected to be below the lower turning point, where the amplitudes are substantially lower, further reducing its importance. 
If core dissipation is the correct interpretation of tidal $Q$ for Jupiter then it probably implies a similar, ``low'' $Q$ (relative to our suggested value) for normal modes, but only for those that have significant amplitude in or near the core. 
This will not apply to current observations of large $n$ (see Figure~\ref{eig-plots}). 
We cannot exclude this but note that it increases the difficulty of explaining the observed normal mode amplitudes. 


\section{Possible physical excitation sources}
This section focuses on possible real excitation sources for Jupiter's seismic normal modes. 
Each of these will be modeled crudely. 
The intent here is not to provide highly accurate detailed descriptions of these excitation mechanisms, but rather to simply test if the general energy scales, timescales, and coupling efficiency expected of them could feasibly be candidates to explain the observed signal. 

\subsection{Turbulent Convection} \label{stochast}
Following the work of P. Kumar (1996) \cite{kumar}, we write the the equation of continuity 
\be \rho' + \nabla \cdot (\rho \mathbf{\xi})=0 \ee
and the acoustic wave equation with a source term 
\be \frac{\partial^2 \rho \xi_i}{\partial t^2}+c^2 \frac{\partial \rho'}{\partial x_i}=\frac{\partial T_{ij}}{\partial x_j} \ee
where 
\be T_{ij}\equiv \rho v_i v_j + p \delta_{ij} -\rho c^2 \delta_{ij} \ee
Combining these equations yields the relationship 
\be \frac{\partial^2 \rho \xi_i}{\partial t^2}-c^2 \nabla^2(\rho \xi_i)=-\frac{\partial T_{ij}}{\partial x_j} \ee
Decomposing displacement into eigenfunctions 
\be \mathbf{\xi}= \sum_{nlm} a_{nlm} \mathbf{\xi}_{nlm} \exp(-i \omega t) \ee
where the amplitudes here are normalized to unit energy according to
\be \omega^2 \int \rho |\xi|^2 dV=1 \ee
Solving produces 
$$ \frac{\partial a_{nlm}}{\partial t} =\frac{- i \omega}{\sqrt{2}} \exp(i \omega t) \int \xi_{q_i} \frac{\partial T_{ij}}{\partial x_j}$$
\be =\frac{i \omega}{\sqrt{2}} \exp(i \omega t) \int \frac{\partial \xi_{{nlm}_i}}{\partial x_j} T_{ij} dV \ee
Following the form of turbulent forcing from Lighthill (1952), $T_{ij}\sim \rho v^2 \delta_{ij}$, we can solve 
\be \frac{\partial A_q}{\partial t} \sim \frac{i \omega}{\sqrt{2}} \exp(i \omega t) \int \rho v^2 \frac{\partial \xi_{{nlm}_r}}{\partial r} dV \ee
So the energy input into the mode $(n,l,m)$ follows the time average amplitude squared 
\begin{equation}
\boxed{\frac{d E_{nlm}}{dt} \sim 2 \pi \omega^2 \int r^2 \rho^2 v_{\omega}^3 h_{\omega}^4 \left[\frac{\partial \xi_{{nlm}_r}}{\partial r} \right]^2 dr}
\label{turbulent-energy}
\end{equation}
where $h_{\omega}$ and $v_{\omega}$ are the turbulent eddies which are resonant with the mode, i.e. they satisfy 
\be \frac{h_\omega}{v_\omega}=\frac{2 \pi}{\omega} \ee
Assuming a Kolomogorov cascade which obeys 
\be v_h = v_H \left( \frac{h}{H} \right)^{1/3} \ee
we have everything needed to solve for the energy input once we solve for $H$ and $v_H$. 
From mixing length theory, we use the planetary length scale for $H$, and we know the convective velocity associated with the large scale motion approximately obeys 
\be v_H \sim 0.1 \left[ \frac{L F_{\text{conv}}}{\rho H_T} \right]^{1/3} \ee
where $H_T$ is the temperature scale height. 
Solving this to an order of magnitude assuming Jupiter's entire flux is available for convective flux, using Jupiter's average density and assuming $L/H_T\sim 10$, we obtain $v_H \sim 3$cm s$^{-1}$. 
Solving for $h_\omega$ and $v_\omega$ give 
\be h_\omega \sim 140\text{cm} \left(\frac{10^{-3}\text{s}^{-1}}{\omega} \right)^{3/2} \ee
\be v_\omega \sim 0.03\text{cm s}^{-1} \left( \frac{10^{-3}\text{s}^{-1}}{\omega} \right)^{1/2} \ee
The Reynold's number for these values is of order $10^2-10^3$, so it should still be above the minimum Kolomogorov microscale. 
Using these values and substituting them into Equation~\ref{turbulent-energy} produces the red curve amplitudes on Figure~\ref{water}. 
The amplitudes are orders of magnitude too small to explain the observed normal mode velocity amplitudes, but it is important to note the qualitative behavior of the amplitude spectrum, which shows most of the power in the lowest frequency modes with relatively diminished power in higher frequency modes. 
It is worth noting that the expected convective velocities increase near the surface, as density rapidly decreases but heat flux remains relatively constant. 
This can increase convective velocities by an order of magnitude over a small distance, which can affect the resultant energy input. 
Such detailed calculations are beyond the scope of this paper, but we note that our simplified calculations returned the expected result that mode amplitudes excited using this mechanism are about three orders of magnitude smaller than on the sun, as we would naively expect based on the order of magnitude arguments from Section 1.

\subsection{Meteor Strikes}
As much of this paper has, the idea of a meteor strike's excitation will closely follow the work of Dombard and Boughn \cite{dandb} for the Shoemaker-Levy/9 Jovian cometary impact. 
Here the primary excitation source is a monopolar explosion, which occurs after the meteor reaches a certain pressure depth. 
Since the explosion happens very quickly, we can approximate it as a Heaviside Theta function so that 
$\phi(t)\rightarrow \Theta(t)$. 
Assuming the comet explodes at the 50 bar level, and taking the energy of the explosion to be $10^{30}$ergs (an optimistic estimate; this corresponds to an upper bound on extremely large impacts like SL9 \cite{dandb} and should be treated as an upper bound), and assuming an impact of this magnitude happens approximately every 50 years, we get negligible equilibrium amplitudes on the order of microns per second. 
If we use smaller impact energies, the excitation is correspondingly smaller. 
We did not bother to include smaller, more frequent impacts in this calculation because as argued above only the most energetic events significantly affect the equilibrium amplitudes. 

\subsection{Storms}
As all models in this paper, the formulation for storm models will be greatly simplified. 
The types of storms we are interested for these purposes form when a parcel of moist air is lifted to the level of free convection (LFC) by some external driving force. 
Once there, some moisture precipitates out of the parcel, releasing latent heat. 
This heat causes the parcel to warm and expand, which causes it to become buoyant and rise. 
As it rises and expands, the parcel cools, allowing more condensation and releasing more latent heat. 
As this moist parcel rises, it will follow a moist adiabat, causing it to be warmer than the surrounding environment at all levels above the LFC. 
The parcel will continue to rise until it equilibrates with its surroundings. 
On Earth, this happens at the inversion layer, or the tropopause. 
This same basic picture applies to water storms on Jupiter \cite{stoker}, with the important difference that on Earth water vapor is less dense than the ambient air, while the opposite is true on Jupiter. 
To model how such a process would affect the surrounding atmosphere, we consider the relevant forces. 
As the parcel rises, it pulls air along with it. 
The characteristic force is the buoyancy of the parcel, so 
\be f_0 \sim \Delta\rho g V \ee
where $V$ is the volume of the parcel and $\Delta \rho$ is the change in density resulting from the release of latent heat, i.e. 
\be  \frac{\Delta \rho}{\rho}=\frac{L_v f}{c_p T} \ee
where $f$ is the mass fraction of the condensing constituent and $L_v$ is the latent heat of vaporization. 
The distance over which this dipole acts would scale with the distance the parcel rises.
substituting these values into the equation for dipole forcing, we obtain 
\be F_0 \sim \frac{E_s \frac{\partial \xi_r}{\partial r}|_{r=r_0}}{\iiint \rho |\xi|^2 dV} \label{thisdip} \ee
where $r_0$ is the height of the cloud deck. 
Now to calculate the appropriate storm energy that couples to the mode. 
If a rising column of air like this were to originate deep within the atmosphere, it could in principle rise all the way to the statosphere. 
However, if it started many order of magnitude higher in pressure, the parcel itself would probably break apart and lose its coherence after about a scale height. 
Alternatively, it could keep rising until it hit a cloud deck above it, providing the lifting needed to lift the parcel in front of it above the LFC, while the droplets that condensed down below have already rained out. 
The dynamics of how such a situation would proceed are complex and uncertain. 
We therefore assume that the height the parcel will rise scales with the environmental scale height  
$ \epsilon \propto H $.\\

The column of rising air will have some characteristic radius $r$ and some height $H$. 
A thin parcel of rising air would then have volume $\pi r^2 dz$, implying a buoyant force of $\pi r^2 \Delta \rho g dz$. 
Each parcel of rising air starts at the cloud deck, and rises a characteristic distance $H$. 
Therefore the work done by each parcel is approximately $\pi r^2 H \Delta \rho g dz$. 
Now integrating over the height of the column, we find the characteristic storm energy from Equation~\ref{thisdip} to be about 
\be E_s \simeq \pi r^2 H^2 \Delta \rho g \ee
The power output by water storms in Jupiter is about 3.3Wm$^{-2}$ \cite{stormdata}, which is a significant fraction of Jupiter's total heat budget. 
The characteristic size of convective columns can be large, on the order of 100km or more. 
If this is the case, the effect of entrainment on column buoyancy is negligible \cite{stoker}. 
When a convective plume rises, it does so by releasing latent heat. 
The total latent heat released by this process is approximately the total mass of condensate in the column 
\be E_L\sim \pi r^2 H \rho f L_v \ee
where $r$ is the radius of the convective column. 
The characteristic timescale between such a column rising, then, is just this energy scale divided by the total power output by storms over the whole of Jupiter's surface. 
This gives us $E_L \sim 1.3 \times 10^{26}\text{erg} \implies \tau_s \sim 65$s, and $E_s \sim 3.6 \times 10^{25}$erg if the height of the column is 50km \cite{stoker}. 
This is compatible with our expectations about observed storm activity on Jupiter.
Finally, we model the storm to be a hat function in time with a timescale that scales with the buoyancy timescale 
\be \Delta t \sim \frac{H}{v} \ee
where 
\be v^2 \approx \frac{L_v f}{c_p T} r g \ee
Following through with the calculation and assuming $Q\sim 10^7$, we obtain the expected normal mode velocity spectrum in Figure~\ref{water}.\\
{
\centering
\includegraphics[scale=.8]{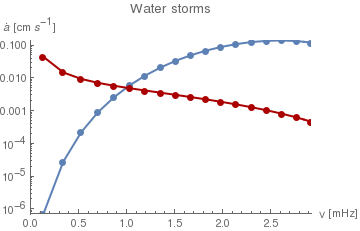}
\captionof{figure}{Amplitude excitation based on estimates for water storm forcing (blue curve). For comparison, the red curve shows the expected amplitude spectrum from stochastic excitation from turbulent convection.}\label{water}
}

Clearly, the amplitudes are orders of magnitude too small to explain the SYMPA data. 
However, the behavior is qualitatively different from the result of turbulent convection; whereas turbulent convection is expected to deposit most energy in low order modes, storm excitation expects more energy in higher order modes. 
This is an important distinction, and these two broad classes of excitation sources can be compared as data at lower frequencies becomes available. \\

However, we have not solved the problem of exciting larger amplitudes than would be expected from turbulent convection. 
Thermodynamically we expect there to be more cloud levels deeper in Jupiter's interior. 
Detailed calculations about the behavior of chemical equilibria and condensation in Jupiter's shallow interior have been carried out by Fegley and Lodders \cite{fegley}, including the posited existence of rock clouds. 
Silicate and iron clouds have been observed on brown dwarfs and posited on hot exoplanets \cite{brown-dwarfs}, and there has even been some modeling of their storm dynamics \cite{lunine}. 
Similar dynamics may well be at play in Jupiter. 
These comparatively refractory species will have much higher latent heats, and can thus be expected to be more energetic than water storms. 
If this were the case, we could follow through the same analysis but assume the length scales $H$ and $r$ used to calculate $E_s$ and $E_L$ is proportional to the relative pressure scale heights between the water cloud deck and the rock cloud deck.
We also substitute the latent heat of vaporization of water ($2.3\times 10^{10}$erg~g$^{-1}$) with the appropriate value for silica ($1.2\times 10^{11}\text{erg}~\text{g}^{-1}$). 
Rock storms must occur deeper in the atmosphere, where pressure, temperature and density are higher. 
We will use parameters at 10kbar in pressure at around 2000K, roughly where we expect silane gas to start producing silica droplets. 
A visualization of this difference is illustrated in Figure~\ref{cartoon}.\\
{
\centering
\includegraphics[scale=.8]{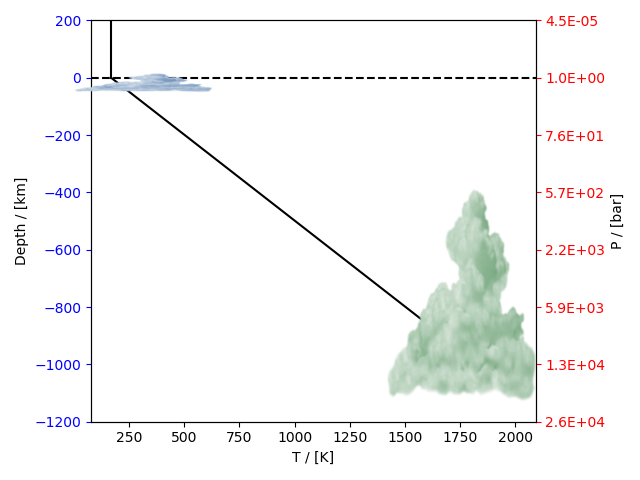}
\captionof{figure}{A cartoon depicting the relative dimensions of water and rock storms. As one dives into the interior, the scale height increases rapidly, which is important for our estimates of storm length scales at these depths. The left y-axis shows depth while the right y-axis shows corresponding pressure. The blue cloud represents the height and location of water storms, while the green cloud represents these same parameters for rock storms.}\label{cartoon}
}
This different depth affects the coupling efficiency for higher frequency modes. 
This is one of several factors which are ignored in Figure~\ref{constraints}. 
The justification for using the latent heat of a silica phase transition as a stand-in for silicate droplet condensation is not immediately obvious, since based on thermodynamic equilibrium chemistry we expect this transition to be a complicated multi-component chemical reaction of silane, iron-carrying vapor, magnesium-carrying vapor, and water vapor to form silicate droplets. 
The dynamics of how such reactions would unfold need future inspection to complete a detailed picture, but for our purposes we are not overly concerned with the details, only the order of magnitude energy scales. 
If we assume the dominant reaction is e.g. silane to silica instead of a silica vapor to liquid phase transition, it affects the outcome by less than 30\%, which is negligible in the context of our order of magnitude consideration. 
Therefore we take a silica phase transition to be a proxy for potentially complicated chemical reactions, noting that the important aspect is the release of heat, not the specific mechanism which causes it. 
As such, we combine the total abundances of silicon, magnesium and iron and take this to be the concentration of silica vapor, in order to simplify the model. 
Finally, we assume that the available energy budget for rock storms is the same as for water storms relative to Jupiter's luminosity. 
Using these parameters and allowing the storm column radius to grow, one can justify using parameters like $E_s\sim 5 \times 10^{31} \implies \tau_s \sim 1.5\times 10^7$. 
Using these parameters, coupling to five kilobar level (the midpoint of the storm on Figure~\ref{cartoon}), the same model produces Figure~\ref{rock}. \\
{
\centering
\includegraphics[scale=.7]{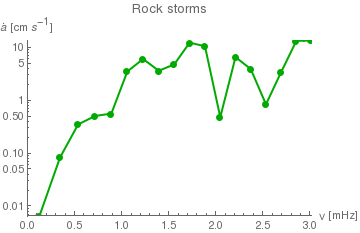}
\captionof{figure}{Amplitude excitation based on preliminary estimates for rock cloud forcing. The non-smooth structure results from the $\sin(\omega\Delta t)$ term. $\Delta t$ here is larger than for water storms, therefore the sinusoid oscillations have a smaller wavelength in frequency space. The specific structure of the curve shouldn't be taken too seriously; the point is the order of magnitude of the velocities which begin to approach the observed values on order of tens of cm/s. }\label{rock}
}
One could easily argue that these parameters are all highly uncertain, and that this is an issue of fine tuning. 
After all, we can adjust the storm parameters to yield any order of magnitude equilibrium mode amplitude we like, in principle. 
But the important point here is not to make an accurate prediction of the behavior of these hypothetical rock storms, whose existence and behavior is largely unconstrained. 
Instead, since we know nothing about rock storms, this analysis is intended to place constraints on the necessary parameters of storm-like activity which could produce the observed equilibrium amplitudes. 
The details of the dynamics of a hypothetical rock storm are highly speculative. 
In this paper we assumed the dynamics were identical to water storms, and just scaled the parameters to their appropriate values accordingly. 
This exercise serves simply to demonstrate an example of a physically plausible mechanism which could excite the observed amplitudes. 

\section{Results and Discussion}
No excitation mechanism investigated here seems to be a clear candidate for producing the observed amplitudes of Jovian seismic modes. 
However, if we are to believe the results, we can place meaningful constraints of the type of source that may cause these observations, and make some predictions about other frequencies based on this. 

\subsection{Excitation source parameter constraints}
The expected turbulent convection is insufficient to explain to observed amplitudes of normal modes. 
Point source excitations, either storms, meteor strikes, or something else, may be able to explain these amplitudes if analyzed more carefully. 
Both monopole and dipole excitation types are of the same form, to first order. 
\be \dot{a_{eq}} \sim \frac{E_s \frac{\partial\xi_r}{\partial r}|_{r=r_0}}{\iiint \rho |\xi|^2 dV} \left( \frac{2 Q}{\tau_s \omega^3} \right)^{1/2} \xi_r(R) \ee
Using this general form, one can place order of magnitude constraints on the necessary bulk parameters needed to excite the amplitudes observed by SYMPA.\\
{
\centering
\includegraphics[scale=0.7]{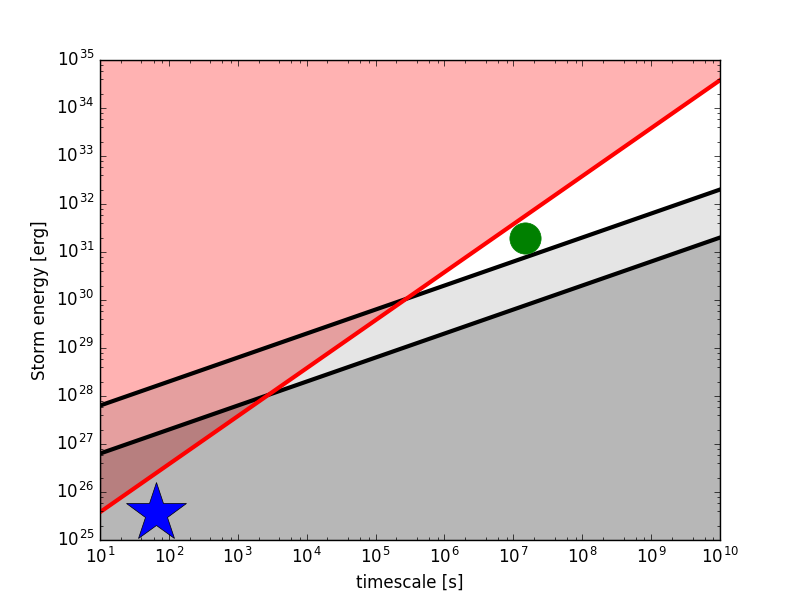}
\captionof{figure}{Assuming a storm-like excitation and holding all other parameters constant, any viable candidate must lie above the black curve in order to explain the Gaulme et. al. results \cite{french}, and below the red curve to satisfy Jupiter's luminosity constraint. The two black curves represent different values of $Q$. The lowest line represents an idealistic $Q=10^8$, above that a more pessimistic $Q=10^6$. 
The blue star represents the excitation from water storms in this parameter space. 
The green point represents the same model scaled to rock clouds. }\label{constraints}
}
Any such mechanism must not violate Jupiter's total energy budget, but must be energetic and frequent enough to excite modes of the observed amplitude in the steady state. 
There is a sliver of parameter space as shown in Figure~\ref{constraints} which could theoretically satisfy these constraints.

\subsection{Predictions for other frequencies}
Using the storm or meteor strike model, or any generic short-lived, localized, stochastic excitation source, we obtain some general features of the power spectrum. 
In particular, low frequencies generated in this way are orders of magnitude smaller than their overtones, since the local gradient of the radial eigenfunction near the surface is much smaller for lower frequencies, and the coupling is therefore weaker. 
In contrast, the red curve on Figure~\ref{water} shows more power in lower frequency modes compared to overtones. 
Future observations which show the power spectrum with better resolution, and in lower frequencies could distinguish between these two basic classes of excitation: global or point source. 

\subsection{Implications for gravity, Juno, Saturn, and ice giants}
Because no unique candidate for excitation has been determined, it's difficult to make predictions for how this may affect Juno's results. 
If the excitation sources are point sources of the sort described in this work, the amplitudes for f-modes, which would most significantly perturb Jupiter's gravity field, would be orders of magnitude smaller than the overtones detected by SYMPA. 
This means that even though the displacement amplitude of normal mode overtones may be on the order of fifty meters, the fundamental modes could self-consistently have displacement amplitudes of mere centimeters. 
The gravity field perturbation caused by the normal modes is still strongest for the lowest frequency modes, since the global coherence of zeroth radial order modes as shown in Figure~\ref{eig-plots} makes them perturb the gravity field much more strongly than oscillatory, higher order modes. \\

We can decompose the gravity field into a sum of gravity harmonics 
\begin{equation}
\Phi(r,\theta,\phi)=\frac{1}{R} \sum_{l=0}^{\infty} \sum_{m=0}^l \left( \frac{R}{r} \right)^{l+1} (C_{lm} \cos(m\phi) + S_{lm} \sin(m\phi) P_l^m(\cos\theta))
\end{equation}
Because both gravity harmonics and normal modes are defined by spherical harmonics, a given normal mode's gravity perturbation can be completely described by a single gravity harmonic term. 
If we wish to ask whether a given normal mode will be detectable, we can compute an illustrative example by considering how $J_2$ is affected by $\xi_{n20}$. 
To calculate this change, we must compute the density perturbation $\delta \rho_{nlm}$ from a displacement eigenfunction $\xi_{nlm}$. 
We can do this simply by using the continuity equation 
\begin{equation}
\delta \rho = \nabla \cdot (\rho \xi)
\label{density}
\end{equation}
The shape of these density eigenfunctions are shown on Figure~\ref{densities}. 
{
\centering
\includegraphics[scale=0.7]{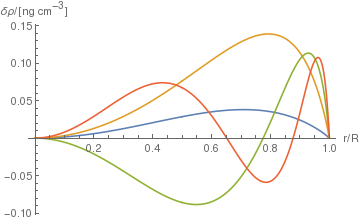}
\captionof{figure}{Normalized density eigenfunctions for the first few l=2 modes. Notice that the n=1 density eigenfunction has no nodes, even though its corresponding displacement eigenfunction has one. This is a simple consequence of Equation~\ref{density}, since the density is the divergence of the displacement.}\label{densities}
}
To calculate the change in $J_l$ associated with mode $\xi_{nl0}$, we use 
\be 
J_l = - \frac{1}{M R^l} \int r'^l P_l(\cos \theta')\delta \rho(\mathbf{r}') d^3 \mathbf{r}'
\ee

Juno's $\Delta J_2$ $3\sigma$ uncertainty for gravity perturbations is about $10^{-8} \Phi_{\jupiter}$ \cite{juno-sup}
, so we can compute the required amplitudes for gravitational detection of normal modes by Juno. 
This is shown on Figure~\ref{jupiter-gravity}. 
{
\centering
\includegraphics[scale=0.7]{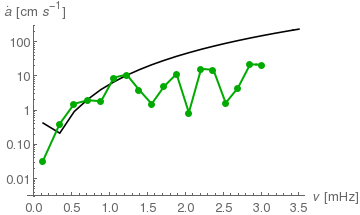}
\captionof{figure}{The black curve represents the 3 sigma sensitivity limit for Juno detecting a variation in $J_2$, and the green curve is identical to Figure~\ref{rock}.}\label{jupiter-gravity}
}
\justifying
Evidently under the assumptions of our model, detection of some normal modes from Juno gravity is plausible. 
However it's right on the edge, and since our results are very imprecise, detection of lack of detection are both plausible outcomes. 

Identical calculations to the ones carried out for Jupiter can be replicated for any planetary model, simply changing input parameters. 
In addition to Jupiter, we have carried out these calculations for Saturn. 
Kronoseismology has developed in a different trajectory from dioseismology, since the seismometers employed for Saturn are the rings themselves. 
Kronoseismology is therefore most sensitive to modes which can resonate with the orbits of ring particles. 
Because there is a gap between the surface of Saturn and the C-ring, only the lowest frequency modes can be detected this way. 
In contrast, dioseismology is performed using time series Doppler imaging, which is most sensitive to the largest velocities and shorter periods, i.e. overtones. 
Jupiter and Saturn are very similar planets, with similar compositions, radii, and heat budgets. 
It is therefore probable that they each behave much more like each other than like stars. 
Turbulent convection as a source of normal mode excitation suffers the same deficiency on Saturn as it does on Jupiter; small convective velocities. 
Convective velocities are on the order of 3cm~s$^{-1}$ for both, much smaller than the sound speed in both cases. 
This would indicate a power spectrum comparable to the red curve on Figure~\ref{water}. 
Amplitudes derived for Saturn's mixed f and g-modes \cite{fuller} do not require additional excitation sources beyond stochastic excitation from turbulent convection to explain \cite{marley1991}\cite{marley1993}. 
For this reason, we must ensure that our storm excitation mechanism, which was used to explain large mode amplitudes on Jupiter, does not produce \textit{excessively large} amplitudes on Saturn. 
In particular, we can compute the mode excitation by observed storms expected based on our model. 
Based on the arguments leading up to Equation~\ref{fundamental}, the water storms on Saturn may be much more important for mode excitation than the water storms of Jupiter. 
While Jupiter has continuous thunderstorms happening all over its surface, Saturn has just one hugely energetic storm every few decades \cite{andy}. 
The most recent Great Storm on Saturn occurred in 2011, and was observed by Cassini, ground based telescopes, and amateur astronomers. 
Similar Great Storms have been seen throughout Saturn's history, occurring on a characteristic timescale of roughly 30 years. 
As demonstrated, this type of excitation (infrequent, large energy) is the most favorable situation to produce high amplitude normal modes. 
The great storm on Saturn releases as much energy as the whole of Saturn does in a year \cite{gws}. 
Assuming $E_s/E_L \sim 10\%$, as is the case for water storms on Jupiter, this provides an approximation for $E_s \sim 4\times 10^{30}$ergs. 
We know events like these occur roughly every 30 years, which directly provides the relevant $\tau_s$. 
We can do a similar analysis to the one applied to Jupiter, but apply parameters relevant to the Saturnian Great Storms and scale our calculated dissipation due to radiative damping to Saturn.
This produces a value of $Q\sim 5\times 10^6$ which is consistent with (although much larger than) the observational lower bound of $Q>10^4$ \cite{kronos}.
Using these inputs we obtain Figure~\ref{saturn-gws}. \\
{
\centering
\includegraphics[scale=.7]{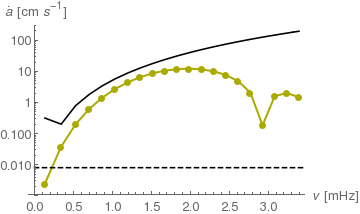}
\captionof{figure}{Saturn velocity amplitudes based on estimates for the Great White Spot 30 year quasi-periodic super storm. The yellow curve represents the expected amplitudes, while the black curve represents the detection limit for Cassini gravity, and the dashed gray line corresponds roughly to Jim Fuller's prediction for f-mode amplitudes on Saturn based on inspection of optical depth variations in the spiral density waves in Saturn's rings raised by its normal modes \cite{fuller}.}\label{saturn-gws}
}
We can use Figure~\ref{saturn-gws} to compare our predictions to expected measurements. 
This calculation did not include dissipation from the core, which could be more important on Saturn than on Jupiter since Saturn's core is known to be relatively large. 
This indicates that for the lowest order modes, storm activity may be comparable in importance to turbulent convection, and that for higher frequency overtones Saturn may have comparable normal mode amplitudes to Jupiter. 
Importantly, the storm excites relatively small amplitudes for Saturn's low order modes. 
If those excitation predictions were too large, it would be evidence against our storm excitation model, since it would be inconsistent with observations. 
Additionally, rock clouds may also play a role in Saturn as they do in Jupiter. 
However, our analysis suggests the Great White Spot alone could theoretically produce p-mode amplitudes on Saturn of the same order as have been observed on Jupiter, an interesting result on its own. 
For this reason we will refrain from further speculation about additional excitation sources. 
Doppler imaging of Saturn may take additional technical advances or dedicated time on larger telescopes, because the light from Saturn that reaches Earth is significantly fainter than that of Jupiter. 
As with Jupiter, it is unclear whether a gravity signal from the normal modes can be expected. 
Certainly additional excitation from rock storms on Saturn could put it over the edge. 
However, stochastic excitation from turbulent convection as we have calculated it certainly cannot produce normal mode amplitudes large enough to produce a gravity signal \cite{marley1991}\cite{marley1993}. 
Therefore if one wishes to invoke normal modes as the explanation for the unexplained component of Saturn's gravity field measured by Cassini \cite{dark-side}, one must consider storms or some other excitation source. \\

In addition to the gas giants, ice giants may prove to be of similar interest for performing planetary seismology from orbit \cite{ice-giants}. 
Three of four multi-billion dollar proposals for missions to either Uranus or Neptune in the coming decades include a doppler imager, which would ideally be capable of detecting seismic normal modes. 
Attempts have been made to measure poseidoseismology (seismology on Neptune) using Kepler K2, although only the reflection of solar oscillations were detected \cite{neptune-k2}. 
Unfortunately, it is difficult to put constraints on what amplitudes to expect without a coherent understanding of the excitation source or an a priori knowledge of the planetary interior. 
Indeed, complicated interactions between the atmosphere and the mantle of the ice giants, immense uncertainty about interior dynamics, general ignorance of the ice giants' bulk interior structure including possible dissipation mechanisms, and universal uncertainty about normal mode excitation theory in giant planets makes constraining the expected normal mode amplitudes exceedingly difficult. 
Rather than attempting a naive quantitative analysis here, we will simply provide some remarks for future work. 
Using an approximation of the equation of state from previous studies of the ice giants' interiors \cite{uranus-interior}, we constructed hypothetical eigenfunctions for Uranus and Neptune which, although highly uncertain, provide an order of magnitude estimate for the general scale of the inertia of these modes and gradients near the surface. 
Uranus and Neptune have much smaller energy fluxes than Saturn and Jupiter, even relative to their total masses. 
Convective velocities should be on the order of 1cm~s$^{-1}$, insufficient to excite amplitudes larger than microns per second. 
However, methane storms have been observed from Earth on Uranus \cite{uranus-storms}, so it is possible that this activity could excite higher amplitude normal mode responses. 
Storm systems observable by telescope are methane storms, but just as rock storms could be at play deeper in Jupiter, water storms could behave similarly deeper in the ice giants. 
Of course, the eventual amplitude depends strongly on the energy and timescales of the storm, as shown in Figure~\ref{constraints}. 
Neptune has a larger luminosity than Uranus, and could therefore in principle produce higher amplitude modes.
It is possible of course that solid phase seismic activity in the mantle could couple very efficiently to the atmosphere to provide higher amplitude responses in the upper atmosphere.
A Uranus quake occurring in a solid phase mantle, for example, could couple efficiently to the dense overlying atmosphere and produce a high amplitude signal in the stratosphere.
Such a mechanism, however, is beyond the scope of this paper. 
Indeed, it is very difficult to place theoretical constraints on ice giant seismic mode amplitudes without making an enormous array of assumptions. 
Since we don't even understand the very basics of ice giant interiors, such assumptions are difficult to defend. 
A more focused effort to characterize normal mode couplings in the ice giants, as well as an elementary understanding of deep moist convection in gaseous interiors, could provide some basic theoretical predictions for normal mode amplitudes for the ice giants, which would be necessary for calibrating a Doppler imager on board a future mission. 
Before such a method could be reliably employed, much further study of giant planet seismology must be carried out, both on the observational and theoretical fronts, as well as further study of ice giant and gas giant interior dynamics.

\section{Conclusion}
The observed amplitudes of normal modes on Jupiter are in great excess of what would be expected based on turbulent convective theory. 
Meteor strikes do not occur frequently enough or with sufficient energy to excite the observed amplitudes either. 
Water storms are extremely frequent, but relatively low energy and with very weak coupling to the normal modes. 
Therefore they cannot come anywhere close to explaining the observed modes. 
The only viable candidate examined in this paper is rock storms. 
It should be mentioned that there are other possible excitation mechanisms not examined in this work that may warrant further study. 
For example, baroclinic instabilities may play a role in seismic mode excitation. 
Additionally, dynamics in the helium rain layer or in a region of deep static stability are potentially worth consideration. 
If the primary excitation source is rock storms, as suggested here, the specific dynamics of the rock storms could significantly affect the outcome. 
In particular, the timescale associated with a rock storm's duration, and the length scales associated with such a storm, might differ significantly from the basic simplifying assumptions presented here. 
However, rock clouds are a promising candidate given the large latent heat of silicates compared to water, as well as the large length scales expected at such a depth with an atmospheric scale height much larger than the upper troposphere. 
Preliminary crude calculations indicate that any storm mechanism invoked to explain the observed amplitudes must occur below the red curve and at least above the lowest black curve on Figure~\ref{constraints}. 
Jupiter may have a rich abundance of storm activity below the visible surface. 
This work suggests this storm activity could feasibly be responsible for the much larger normal mode amplitudes seen on Jupiter compared to predictions. 
More sophisticated models of storm activity may show better coupling between storms and normal modes than we estimated here, which could make these storms a candidate to explain Jupiter's normal modes. 
Similar storms and large scale convection may excite normal modes on the ice giants in a similar fashion, and this topic warrants further study.


\end{document}